\documentclass[12pt]{article}
\usepackage{amsmath,amsfonts,amssymb}
\usepackage{graphics}
\usepackage{lscape}
\usepackage{epsf}

\newcommand{\half}{{\textstyle{\frac{1}{2}}}}
\newcommand{\thalf}{{\textstyle{\frac{3}{2}}}}

 \newcommand{\be}{\begin{equation}}
\newcommand{\ee}{\end{equation}}
\newcommand{\bea}{\begin{eqnarray}}
\newcommand{\eea}{\end{eqnarray}}

\newcommand{\jn}{\Delta{J_{N}}}
\newcommand{\ji}{\Delta{J_{i}}}

\newcommand{\ai}{A_{i}}
\newcommand{\ait}{A_{i}^{T}}
\newcommand{\aik}{A_{i,k}}

\newcommand{\aikt}{A_{i,k}^{T}}

\newcommand{\jp}{J_{+}}
\newcommand{\jm}{J_{-}}
\newcommand{\s}[3]{\sum_{#1=#2}^{#3}}

\newcommand{\upa}{\uparrow}
\newcommand{\doa}{\downarrow}

\title{Mutation model for oligonucleotides fitting a Yule distribution}

\author{C. Minichini\thanks{E-mail: {\ttfamily minichini@na.infn.it}} 
\and A. Sciarrino\thanks{E-mail: {\ttfamily sciarrino@na.infn.it}}}
\date{}

\begin{document}

\begin{titlepage} 

\maketitle
{\itshape{Dipartimento di Scienze Fisiche, Universit{\`a} di Napoli
``Federico II''
and I.N.F.N., Sezione di Napoli -
Complesso Universitario di Monte S. Angelo
Via Cintia, I-80126 Naples, Italy}}\\

\begin{abstract}
 A spin chain, describing a nucleotides sequence, is identified by the the labels of a  vector 
state of an 
 irreducible representation of $\mathcal{U}_{q \to 0}(sl_2)$. A 
 master equation for the distribuion function is written, where the 
intensity of 
 the one-spin flip is assumed to depend from the variation of the 
labels. The 
   numerically computed equilibrium distribution is  nicely fitted by a Yule
   distribution, 
which is the observed distribution
 of the ranked short oligonucleotides frequency in DNA
 \end{abstract}

\vfill
\begin{flushright}
preprint nr. DSF-41/2004
\end{flushright}

\end{titlepage}

\section{Introduction}

Spin chains, both in the classical and quantum version, are extremely important tools to understand 
various physical situations and, in some cases, provide completely 
soluble models. An interesting field of applications of these models 
is 
the theory of molecular biological evolution.  Since 1986, when
Leuth$\ddot{a}$usser \cite{Leu86} put a correspondence between the 
Eigen model
  of evolution \cite{Eigen} and a two-dimensional Ising model, many 
  articles have been written representing biological systems as  
  spin models. Recently in \cite{BBW} it 
has been shown that the parallel mutation-selection model can be put 
in 
correspondence with the hamiltonian of an Ising quantum chain and in
\cite{Saakian} the Eigen model   has been 
mapped into the hamiltonian of one-dimensional quantum spin chains.
 In this approach the genetic sequence is specified by a
sequence of spin values $ \pm 1$.  DNA is build up a sequence of a 
four basis 
or nucleotides which are usually identified by their letter: C, G, T, 
A (T being replaced by
 U in RNA), C and U  (G and A) belonging to the purine family, 
 denoted by R (respectively to the pyrimidine family, denoted by Y).
 Therefore in the case of
 genome sequences each point in the 
 sequence
 should be identified by an element of a four letter alphabet. 
  As a simplification one identifies each 
 element according to the  purine or pyrimidine nature, reducing to a 
 binary value, see \cite{HWB} for a four-state quantum chain approach.
 As standard  assumption,   the strength 
of the mutation is assumed to depend from the Hamming distance 
between two sequences, i.e. the number of sites with different 
labels.
  Moreover usually it is assumed that the mutation matrix 
elements are vanishing for Hamming distances larger than 1, i.e. for 
more than one nucleotide changes. 
 The main aim of the works using this approach, see \cite{WBG,BW,HRW},
is to find, in different landscapes, the mean ``fitness" and the ``biological surplus", in the framework of biological population 
evolution. 
At our knowledge these has been no attempt to apply spin models to 
obtain the observed equilibrium distribution of oligonucleotides 
in DNA. Martindale and Konopka \cite{MK}, indeed, have remarked that 
the ranked short (ranging from 3 to 10 
nucleotides) oligonucleotide frequencies, in both coding and 
non-coding region of DNA, follow a Yule 
distribution. We recall that a Yule distribution is  given by $f = a 
\, {n^k}\, {b^n}$,
 where $n$ is the rank and $a$, $ k < 0$ and $b$ are 3 real parameters.
 In order  to face this problem, in this paper we propose a  
 spin model   in which the intensity of the 
transition matrix depends in some way from the whole distribution of 
the 
nucleotides in the sequence. At present we assume that the transition 
matrix 
does not vanish only for total spin flip equal $\pm 1$, 
 
\section{Mutations and Crystal basis}

A sequence of N ordered nucleotides, characterized only by the purine 
or 
pyrimidine character, that is a string of $N$ binary labels or spin, can be represented as a vector belonging to the 
N-fold tensor product of the fundamental irreducible representation 
(irrep.) (labelled by $ J =1/2$) of $\mathcal{U}_{q \to 0}(sl_2)$ 
\cite{Kashi}, which is usually called crystal basis representation. 
This parametrization allows to represent, in a simple way,
the mutation of a  sequence as a linear transformation between vectors, which
can be subjected to selection rules and  
whose strength  depends from the two concerned states. 
So we identify a N-nucleotidic sequence as a vector
\be
 \mid \mathbf{J} \rangle = \mid J_3, J^N, \ldots,
	  J^{i}, \ldots, J^2 \rangle
	\label{eq:def}
\ee 
where  $\; J^{N} \; $ labels the irrep. which the vector belongs to,
$\; J_{3} \; $ is the value of the 3rd diagonal generator of
 $\mathcal{U}_{q \to 0}(sl_2)$ ($2J_{3} = n_{R} - n_{Y}$, $n_{x}$ 
being the number of $x$ elements in the sequence)
and $\; J^{i} \; $ ($ 2 \leq i \leq N - 1$) are $N-2$ labels needed
to remove the degeneracy of the irreps. in the tensor product,  in 
order to 
completely identify the state and which can be seen as identifying 
the irrep.
 which the sequence truncated to the $i$-th element belongs to.
 As an example, we  consider a trinucleotidic string
($N=3$) and label the eight different spin chains in the following 
way($\mid {J_3},{J^3},{J^2}\rangle$,
$R \equiv \half \equiv \upa, \; Y \equiv -\half \equiv \doa$): 
\begin{eqnarray*}
\upa\doa\doa &=& \mid -\half,\half,0 \rangle  \;\;\;\;
\upa\doa\upa = \mid \half,\half,0 \rangle \\
\doa\upa\doa &=& \mid -\half,\half,1 \rangle  \;\;\;\;
\upa\upa\doa = \mid \half,\half,1 \rangle \\
\doa\doa\doa &=& \mid -\thalf,\thalf,1 \rangle  \;\;\;\;
\doa\doa\upa = \mid -\half,\thalf,1 \rangle \\
\doa\upa\upa &=& \mid \half,\thalf,1 \rangle  \;\;\;\;\;\;\,
\upa\upa\upa = \mid \thalf,\thalf,1 \rangle.
\end{eqnarray*} 
At this stage the crystal basis  provides an alternative way of labelling any 
finite spin sequence, mapping any sequence in a vector state of an 
irrep., but we know that in physics and mathematics the 
choice of appropriate variables is of primary importance to face a 
problem. Indeed we argue that these variables are suitable to 
partially describe non local events which affect the mutations.
Flipping the total spin by $\pm 1$ ($\Delta J_3= \pm 1 $) can induce 
a transition 
to a vector belonging or not 
belonging to the irrep. of the original sequence. One can easily realize 
that  to identify a nucleotidic sequence as a vector of 
an irrep. requires to fix the number of RY ``contracted couples" 
occurring in the considered sequence
(contraction should be 
 understood in the same sense of contraction of creation-annihilation 
 operators in the Wick expansion). 
   Therefore flipping a spin implies or the creation   or the 
deletion 
  of 
   a RY contracted couple, corresponding respectively to a variation 
   of -1 o +1 of the value  of the $J^N$   and, possibly, of some 
others $J^i$
($2 \leq i \leq N-1$), or to leave unmodified the number of 
contracted couples (so that   $\Delta{J^{N}}=0$, but 
 $\Delta{J^{i}} \neq 0$ for some values of i). Note that 
alternatively one can
identify $1/2 \equiv (C,G)$ and $-1/2 \equiv (T,A)$. Below we give 
phenomenological
arguments for our choice.

\section{Transition operators}

Let us consider a N-nucleotidic string and classify the different 
transitions on the string labels $J_{3}, J^{N},\ldots,J^{2}$. We can 
distinguish different string configurations around the $i$-th 
position,
so that a single nucleotide mutation in $i$-th position can 
correspond to different variations in the string labels. We call 
\textit{left}
(\textit{right}) \textit{side free}  the nucleotides
on the left (right) of $i$-th position and not contracted    with
another one on the same side. 
Let $R_l$ be the initial (before mutation) number of the \textit{left 
side free} purines and $Y_r$
the initial number of the \textit{right side free} pyrimidines.
We want to count the total number of contracted $RY$ couples (before 
and after
mutation) in the
string, so we call $R_{in}$ ($R_{fi}$) the number, in the initial 
(final) state,
of $R$  preceding
some $Y$ and not contracted with any $Y$ on their side. In the same 
way, with $Y_{in}$ ($Y_{fi}$)
we refer to the number of $Y$ following some $R$ and not contracted 
with any $R$ on their side.
If a $R \rightarrow Y$ mutation ($\Delta{J_3}=-1$) occurs in $i$-th 
position, then $R_{in}=R_{fi}+1$
and $Y_{in}=Y_{fi}-1$, where $R_{in}=R_{l}+1$ and $Y_{in}=Y_{r}$. 
 We are interested in finding the stationary or equilibrium configuration of the $2^{N}$
different possible 
 sequence. Writing $p_{\mathbf{J}}(t)$ the probability distribution at 
 time $t$ of the sequence identified by the vector $\mid \mathbf{J} 
 \rangle$, a decoupled version of selection mutation equation (see
 \cite{Hof} for an exhaustive review), for a haploid organism, can be written as
 \be
 \frac{d}{dt} \, p_{\mathbf{J}}(t) =  p_{\mathbf{J}}(t)\left(R_{\mathbf{J}}-
\sum_{\mathbf{K}}\;R_{\mathbf{K}}\; p_{\mathbf{K}}(t)\right)+\sum_{\mathbf{K}}\;M_{\mathbf{J,K}}\; p_{\mathbf{K}}(t)
 \label{eq:ME}
\ee 
where $R_{\mathbf{K}}$ is the Malthusian fitness of the sequence corresponding to
the vector $\mid \mathbf{K} \rangle$ and $;M_{\mathbf{J,K}}$ are the entries of a
mutation matrix $M$ which satisfies 
\be
{M}_{\mathbf{J},\mathbf{J}} = - \, \sum_{\mathbf{K} \neq 
\mathbf{J}} \; {M}_{\mathbf{J},\mathbf{K}}
\label{eq:norm}
\ee
The equation (\ref{eq:ME}) is reduced to
\be
\frac{d}{dt} \, x_{\mathbf{J}}(t) = \sum_{\mathbf{K}}\;\left(H+M\right)_{\mathbf{J,K}}
\;x_{\mathbf{K}}(t)
\label{eq:Hamilt}
\ee
where
\be
x_{\mathbf{J}}(t) =  p_{\mathbf{J}}(t)\exp\left(\sum_{\mathbf{K}}\;R_{\mathbf{K}}\;
\int_{0}^{t}\; p_{\mathbf{K}}(\tau)\,d\tau \right)
\label{eq:transf}
\ee
and $H$ is a diagonal matrix, with fitness as entries.

In our model the mutation matrix is written as the sum of the 
following partial mutation matrices (T means transposition):

\begin{itemize}

\item If $\mathbf{R_l}=\mathbf{Y_r}$ 
 we distinguish two subcases:

\begin{enumerate}

\item $R_{l}=Y_{r}\neq{0}$
 \be
{M_1}=\s{i}{2}{N-1}\s{k}{i+1}{N} \, \alpha_{1}^{ik} \,( {\aik\jm + 
\jp\aikt})
\label{eq:1}
\ee
 \item $R_{l}=Y_{r}={0}$
\be
{M_2}=  \alpha_{2} \, (\jm+\jp) \label{eq:2}
\ee
\end{enumerate}
\item If $\mathbf{R_l}>\mathbf{Y_r}$, 
we distinguish two subcases:
\begin{enumerate}
\item $Y_{r}=0$ 
\be
{M_3}=\s{i}{2}{N} \, \alpha_{3}^{i} \, ({\ai\jm + \jp\ait}) 
\label{eq:3}
\ee
\item $Y_{r}\neq{0}$
 \be
{M_4}=\s{i}{3}{N-1} \, \alpha_{4}^{i} \, ({\ai\jm + \jp\ait}) 
\label{eq:4}
\ee
\end{enumerate}
\item If $\mathbf{R_l}<\mathbf{Y_r}$,
 we distinguish two subcases:
\begin{enumerate}
\item $R_{l}=0$
 
 \be
{M_5}=\s{m}{2}{N} \, \alpha_{5}^{m} \,({\jm A_{m}^{T} + 
A_{m}\jp}) 
\label{eq:5}
\ee
\item $R_{l}\neq0$
 ($2 \leq N-2; i+1 \leq k \leq N-1$).
\be
{M_6}=\s{i}{2}{N-2}\s{k}{i+1}{N-1} \, \alpha_{6}^{ik} \,
({\aik \jm A_{k+1}^{T} + \aikt A_{k+1} \jp})  \label{eq:6}
\ee
\end{enumerate}
\end{itemize} 
where  $\jp$ and $\jm$ are the \textit{step operators} (to save space 
we do not write their explicit form) defined by 
Kashiwara
\cite{Kashi}, acting on the states of an irrep.
 with highest weight $J^{N}$,  i.e. 
inducing 
a mutation $\ji=0, \; \forall i \neq N$,
\be
J_{\pm} \mid \mathbf{J} \rangle = \mid J_3 \pm 1, J^N,..,
	J^k,.., J^2 \rangle 
\ee
and 
\bea
A_{i,k} \mid \mathbf{J} \rangle &=& \mid J_3, J^N,..,
	J^k, J^{k-1}-1,.., J^{i}-1, J^{i-1},.., J^2 \rangle 
	\nonumber  \\     
& & (2 \leq i \leq N-1 \;\;\;\; i+1 \leq k \leq N) \label{eq:ik} 
\eea
\bea
A_i \mid \mathbf{J} \rangle &=& \mid J_3, J^{N}-1, \ldots,
    J^{i}-1, J^{i-1}, \ldots, J^2 \rangle
    \nonumber    \\
& & (2 \leq i \leq N) \label{eq:i} 
\eea
Therefore $\aikt$ is the operator which connects vectors differing by +1 
in the value of $J^l$, for $k-1 \leq l \leq i$.
A few words to comment on the above equations. Let us consider a 
mutation $R \rightarrow Y$, which involves a transition
$\jn=-1$ (case $R_{l}>Y_{r}$) and
entails $\Delta{J_3}=-1$, we
have to apply the operator $\jm$, as well as the operator $\ai$. Of 
course, first we have to lower by
1 the value of $J_3$, then to modify $J^N$, otherwise the initial 
state may 
eventually be annihilated, even if the
transition is allowed (in the case $J^{N}-1<J_3$).
Likewise, in corrispondence of a transition $Y \rightarrow R$ 
($\Delta{J_3}=+1$), first the change
$J^{N} \rightarrow J^{N}+1$ has to be performed, then $J_3 \rightarrow 
J_{3}+1$.
Clearly in eq.(\ref{eq:ME}), assuming equal rates for mutations and for 
back mutations, we have to sum the 
operator, which gives rise to the
transition $Y \rightarrow R$ with that one which leads to $R \rightarrow Y$, that is
\be
{A_{i}}J_-  + {J_+}{A_{i}^{T}}
\ee
This operator leads to the mutation $Y \rightarrow R$ or $R 
\rightarrow Y$ for a nucleotide in
$i$-th position, in a string with $R_{l}>Y_{r}$.
If the mutation $R \rightarrow Y$ corresponds to a rising of $J^{N}$ 
(i.e. a transition with
$\jn=1, \Delta{J_3}=-1$, case $R_{l}<Y_{r}$), first $J^{N}$ has to be 
modified, then $J_3$;
therefore we write  
\be
  {J_-}{A_{m}^{T}} + {A_m}J_+
\ee
 The above operator gives rise to mutations $R \rightarrow Y$ and $Y 
\rightarrow R$
for a nucleotide in $i$-th position, preceding the $m$-th one, in the 
case $R_{l}=0, Y_{r}\neq{0}$.
Let us remark that eq.(\ref{eq:4}) is included in  
eq.(\ref{eq:3}), if the coupling constants $\alpha$ are assumed equal;
 in  eq.(\ref{eq:6}), only  the writing order for  $A_{k+1}$ (and its 
transposed)
and $J_{\pm}$ has to be respected.
 Assuming now that the coupling constants do not depend on $i,k,m$, 
we can write the mutation matrix $M$ as 
 \be
{M}=\mu_{1}({M_3}+{M_5})+\mu_{2}{M_1}+\mu_{3}{M_2}+\mu_{4}{M_6} + M_{D}
\ee
where $M_{D}$ is the diagonal part of the mutation matrix defined by 
eq.(\ref{eq:norm}).
 The scale of the values of the 
coupling constants of $M$ is suggested by the phenomenogy. We want to write 
an interaction
term which makes the mutation in alternating purinic/pyrimidinic 
tracts less likely than
polipurinic or polipyrimidinic ones. We mean as a single nucleotide 
mutation in a polipurinic
(polipyrimidinic) tract, a mutation \emph{inside} a string with all 
nucleotides $R$ ($Y$), i.e. 
a  highest (lowest) weight state. Such a transition corresponds to the 
selection rules
$\jn=-1$,$\Delta{J_3}=\pm1$, i.e. a transition generated by the 
action of $M_3$ and $M_5$.
In the mutation matrix $M$, we give them a coupling constant 
smaller than the 
We introduce, for  $ \; \Delta J_{3} = \pm 1 \;$, only four
 different mutation parameters $\;\mu_{i}\;$ ($i = 1,2,3,4$), with
${\mu_1}<{\mu_k} \; k>1$.
\begin{enumerate}
\item $\mu_{1}$ for mutations which change the irrep., $ \; \Delta 
{J^N} = \pm 
1 \; $, and include the spin flip inside a highest or lowest weight 
vector; 
\item $\mu_{2}$ for mutations which do not change the irrep., $ \; 
\Delta {J^N} = 0 
 \; $,  but modify other values of $J^{k}$,  $\; \Delta J^{k} = \pm 
1$ ($2 \leq k \leq N-1$);
\item $\mu_{3}$ for mutations which do neither change the irrep., $ \; 
\Delta J^{N} = 0$,
      nor the other values of $J^{k}$, $\; \Delta J^{k} = 0$, ($2 
\leq k \leq N-1$);  
\item $\mu_{4}$ for mutations which change the irrep., $ \; \Delta J 
= \pm 
1 \; $, but only in a string with $0 \neq R_{l} < Y_{r}$.
\end{enumerate}
 We do not introduce another parameter, for mutations
generated by $M_4$, i.e. $i$-th nucleotide mutation in a string with
$R_{l} > Y_{r} \neq 0$, to not distinguish, in a polipurinic string, 
  a mutation according to its position.
   
\section{Results}

\begin{figure}[tbh]
\begin{center}
\framebox{\epsfxsize=0.6\textwidth
\epsffile{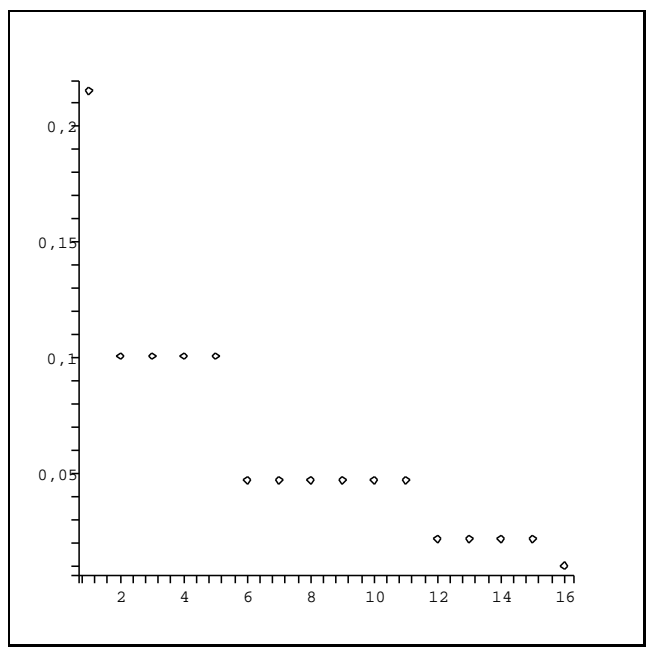}}
\end{center}
\caption{Rank ordered distribution of equilibrium population (N=4) obtained for an Hamming 
transition matrix, with 
$\beta=0.60$.}
\label{stepH16}
 \end{figure}

\begin{figure}[tbh]
\begin{center}
\framebox{\epsfxsize=0.6\textwidth
\epsffile{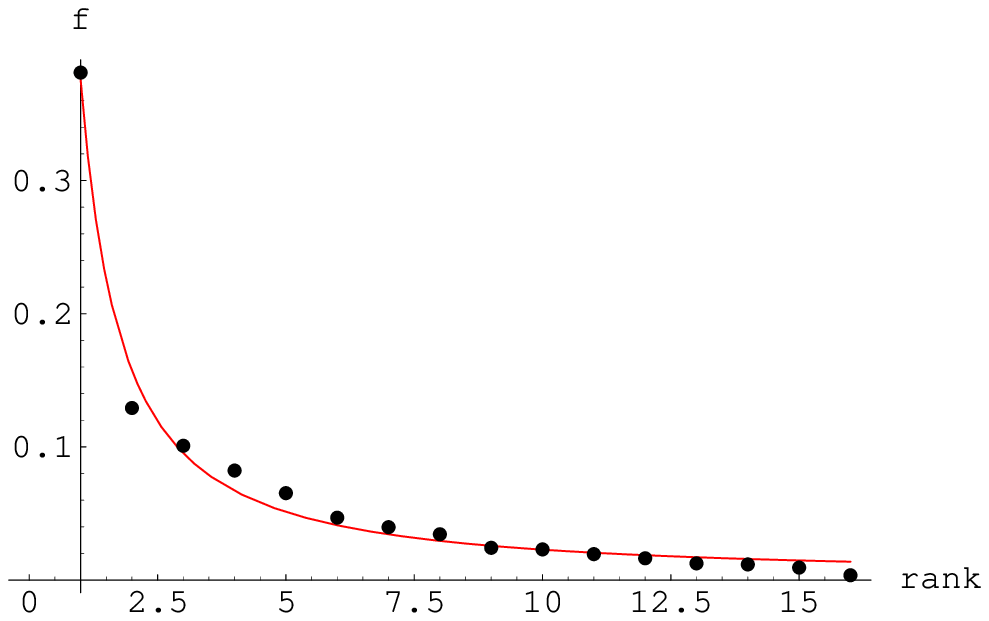}}
\end{center}
\caption{Rank ordered distribution of equilibrium population (N=4) for a 
transition matrix $M$ with 
$\epsilon=0.25,\;\gamma=\delta=\eta=0.50$.
The distribution was fitted by a Yule function (continuous line) 
$f=a{R^k}{b^R}$ ($R$ is the
rank).The parameters was estimated as $a=0.37,\,b=1.02,\,k=-1.28$.}
\label{YuleH16}
 \end{figure}

\begin{figure}[tbh]
\begin{center}
\framebox{\epsfxsize=0.6\textwidth
\epsffile{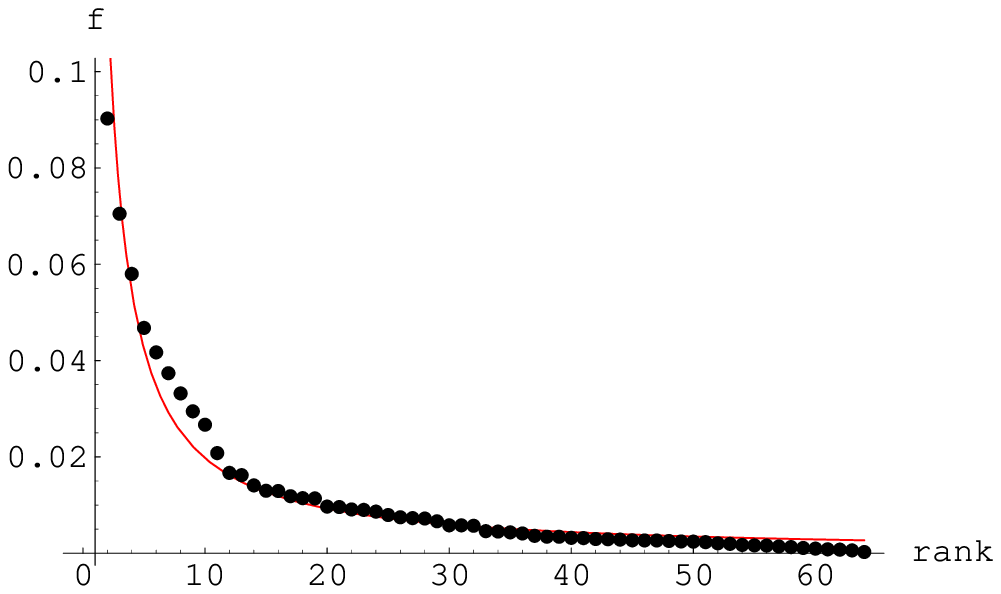}}
\end{center}
\caption{Rank ordered distribution of equilibrium population (N=6) for a transition matrix 
$M$, with
$\epsilon=0.25,\;\gamma=\delta=\eta=0.5O$. The distribution was fitted 
by a Yule
function (continuous line) $f=a{R^k}{b^R}$ ($R$ is the rank). The 
parameters was estimated as $a=0.26,\,b=1.00\,k=-1.11$.} 
\label{N=6}
 \end{figure}

The  evolution equation of the model for the probabilities will be written in 
terms of the matrix  $\bar{H}= H + M + \lambda\mathbf{1}$,
where the fitness can be $H=J_3$ (purely additive fitness)
 and $\lambda$ is choosen in such a way to guarantee $\bar{H}$ is
positive. Being $H + M$ irreducible, the composition of equilibrium population is
given by
\be
p_{\mathbf{J}} = 
\frac{\tilde{x}_{\mathbf{J}}}{\sum_{\mathbf{K}}\;\tilde{x}_{\mathbf{K}}}
\ee
where $\tilde{x}_{\mathbf{J}}$ is the Perron-Frobenius eigenvector\cite{EDM} of
$\bar{H}$. 
 We 
look for a numerical solution, with a suitable choice of the value of 
the parameters, for N = 3,4,6.
Before solving numerically the model, we point out
 explicitly its main features. 
 $M$
 describes an interaction on the $i$-th spin neither depending on the
 position nor on the nature of the closest neighbours, but which  
takes into account, at least partially, the effects on the transition 
in the $i$-th site of the 
distribution of all the spins, that is non local effects. 
 Indeed it
 depends on the ``ordered" spin orientation surplus on the left and on
 the right of the $i$-th position. Should it not depend on the order, 
it
 may be considered as a mean-field like effect. Moreover $\Delta J_3 =
 \pm 1$ transitions   are allowed, which, e.g. for N = 4, can be considered or as the
 flip of a spin combined with an exchange of the two, oppositely
 oriented, previous or following spins or as the collective flip of
 particular three spin systems, containing a two spin system with
opposite spin orientations (see example below).
Biologically, the transition depends in some way on the  "ordered"
 purine surplus on the left and on the right of the mutant position.
 Let us briefly comment on the physical-biological meaning of the
 ``ordered" spin sequence. Our aim is to study finite oligonucleotide 
 sequence in which a beginning and an end are defined. This implies we 
 can neither make a thermodynamic limit on $N$ nor define periodic 
 conditions on the spin chain. So we have to take into account the 
 ``edge" or ``boundary" conditions on the finite sequence. An analogous 
 problem appears in determing thermodynamic properties of short 
 oligomers and, in this framework, in \cite{GB} the 
 concept of fictitious nucleotide pairs E and E' has been introduced, 
 in order to mimick the edge effects. The ordered couple of  RY takes 
 into account in some way the different interactions of R and Y with 
 the edges.
  For example, the transition matrix, on the above basis (for N = 3) 
is the following
one, up to a multiplicative dimensional factor $\mu_{0}$
\begin{equation}
\label{matrixModel}
M = 
\left(
\begin{array}{cccccccc}
x & \delta & 0 & \gamma & \epsilon & 0 & \epsilon & 0 \\
\delta & x & 0 & 0 & 0 & \epsilon & 0 & \epsilon \\
0 & 0 & x & \delta & \epsilon & 0 & \epsilon & 0 \\
\gamma & 0 & \delta & x & 0 & \epsilon & 0 & \epsilon \\
\epsilon & 0 & \epsilon & 0 & x & \delta & 0 & 0 \\
0 & \epsilon & 0 & \epsilon & \delta & x & \delta & 0 \\
\epsilon & 0 & \epsilon & 0 & 0 & \delta & x & \delta  \\
0 & \epsilon & 0 & \epsilon & 0 & 0 & \delta & x
\end{array}
\right)  
\end{equation}
where the diagonal entries are not explicitly written, and are given by (\ref{eq:norm}).
Note that the above matrix depends only on three coupling 
constants
due to the very short length of the chain. For $N \ge 4$ the 4th 
coupling constant
(denoted in the following by $\eta$) will appear.
Let us emphasize that the mutation matrix $M$ (\ref{matrixModel}) does not 
only
connect states at unitary Hamming distance. As an example, we write 
explicitly the
transitions from $\mid \half,\half,0\rangle$ ($\upa\doa\upa$) 
and from
$\mid -\half,\half,0\rangle$ ($\upa\doa\doa$)
\begin{eqnarray*}
\upa\doa\upa \longrightarrow
\left\{
\begin{array}{c}
\upa\upa\upa \\
\doa\doa\upa \\
\upa\doa\doa
\end{array}
\right.& \qquad \upa\doa\doa \longrightarrow
\left\{
\begin{array}{c}
\doa\upa\upa \\
\doa\doa\doa \\
\upa\upa\doa \\
\upa\doa\upa
\end{array}
\right.
\end{eqnarray*}
 For lack of space, we 
do not explicitly write the 
 mutation matrix, which  allows transitions only between chains at
Hamming distance equal to one, with coupling constant $\beta$. Notice 
howevere that such a (Hamming) matrix is 
not obtained by  eq.(\ref{matrixModel}) putting  $ \delta  = \gamma = 
\epsilon = \beta$. 
 If we order (in a decreasing way) the equilibrium probabilities, we obtain,
using the mutation matrix with Hamming distance, 
a rank
ordered distribution of transition probability like that in 
fig.\ref{stepH16} for $N=4$. Its 
shape does not depend on the value of $\beta$.
  The rank-ordered distribution
of the probabilities shows a plateaux structure: 
every
plateaux contains spin sequences at the same Hamming distance from 
the  sequence with the highest value of the fitness.
Using the mutation matrix  (\ref{matrixModel}), the 
rank ordered probabilities distribution does not show a 
plateaux structure,
but its shape is well fitted by a Yule distribution 
(fig.\ref{YuleH16}), like the observed frequency  
distribution of oligonucleotidic in the strings of nucleic 
acids \cite{MK}.  
Let us observe that we obtain a Yule distribution (and not a plateaux 
structure)
even if all parameters in (\ref{matrixModel}) are tuned at the same 
value, which means that the distribution is the outcome of the model 
and not of the choice of the values of the coupling constants.
Analogous resultes are obtained for $N=6$ 
(fig.\ref{N=6}).
 Let us point out that:

i) our model is not equivalent to a model where
  the intensity  depends on the site  undergoing the 
  transition, or from the nature of the closest neighbours or the 
  number of the $R$ and $Y$ labels of the sequence; indeed 
essentially 
  the intensity depends on distribution in the sequence of 
  the $R$ and $Y$;
  
ii) the ranked distribution of the 
probabilities  follows  a Yule distribution law, but as the value of 
the parameter b is close to the unity, 
  the distribution is equally well 
fitted by a Zipf law ($f = a 
\, {n^k}$), in 
agreement with the remark of \cite{MK}.

\begin{figure}[tbh]
\begin{center}
\framebox{\epsfxsize=0.6\textwidth
\epsffile{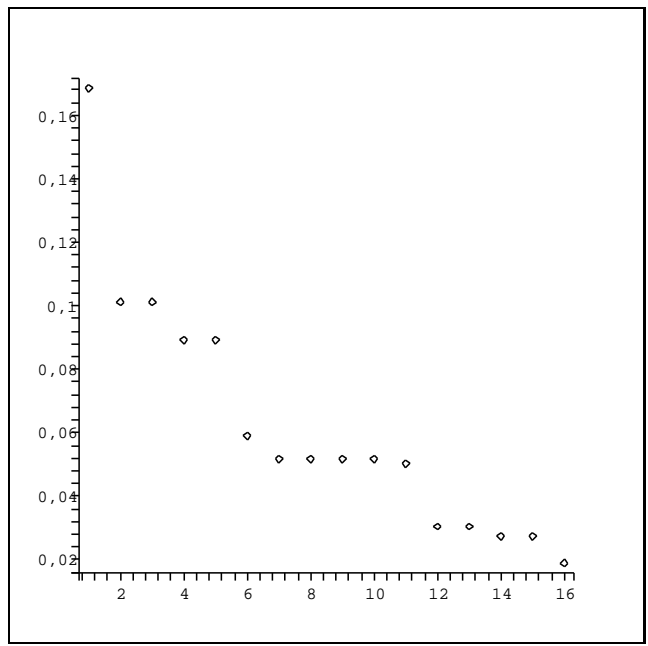}}
\end{center}
\caption{Rank ordered distribution of equilibrium population (N=4) obtained for a
transition matrix allowing the 
 same number of mutations as $M$, between sequences at different Hamming distances} 
\label{fig:cdc}
 \end{figure}

In conclusion we are far from claiming that our simple model is the
only model able to explain 
the observed oligonucleotides distribution, 
for several obvious reasons, but that the standard approach using the 
Hamming distance does not give a Yule or Zipf distribution.
 One may correctly argue that the comparison between the Hamming model, 
depending on only one parameter and taking into account only one 
site spin flip, with our model, which depends on four parameters and 
takes into account spin flip of more than one site, is not meaningful. So we 
have computed the stationary distribution with a mutation matrix 
 not vanishing for Hamming distance larger than one and allowing the 
 same number of mutations as our model. The 
result reported in fig.\ref{fig:cdc} shows that the plateaux structure is 
always the dominant feature. Let us comment on the non point mutations which 
naturally are present in our model. In literature there is 
an increasing number of paper that, on the basis of more accurate data, 
question both the assumptions that mutations occur as single nucleotide
and as independent point event. In a quite recent paper 
Whelan and Goldman
\cite{WG} have presented a model allowing for single-nucleotide, 
doublet and triplet mutation, finding that the model provides 
statistically significant improvements in fits with protein coding 
sequences. We note that the triplet mutations, for which there is no known
inducing mechanism, but which can possibly be explained  by large 
scale event,  called sequence inversion
in \cite{WG}, are indeed the kind of mutations, above discussed, that 
our model naturally describes.
 Doublet mutations do not appear, due to the assumed spin flip equal $\pm 1$, but
 on 
one side some of these mutations are hidden by the binary 
approximation, and on the other side the parameter  ruling such 
mutations, as computed in \cite{WG}, is lower than the one ruling the triplet mutation.
 In conclusion the Hamming distance does not seem
 a suitable measure of the distance in the space of the biological sequences,
 the crystal basis, on the contrary, seems  a better candidate to parametrize
the elements of such space.
  Our model makes use of this parametrisation, allows to modelise
 some non point mutations and exhibits intriguing  and interesting features, 
hinting in the right direction, worthwhile to be further 
investigated.
  In the present simple  version, the model depends 
only on 4 parameters for any N, which are, very likely, not enough to 
describe sequences longer that the considered ones. However the model is rather
flexible and, besides the obvious introduction of 
more coupling constants, allows, e.g., to analyse part of the sequences containing 
hot spots in the mutation (using fictitious edge nucleotides), to 
take into account doublet mutations (indeed the 
operator eq.(\ref{eq:ik}) or $A_{i,i+1}^{T}$ describes a doublet spin 
flip at position i,i+1) and an easy generalisation to four letter 
alphabet.
Although the very short chain, which we were interested in, can be studied numerically 
without any use of the crystal basis,   we propose a 
general algorithm, which can be applied to chains of arbitrary 
length and which can be easily implemented in computers.

\end{document}